\documentclass[10pt,twocolumn]{IEEEtran}
\usepackage{amssymb}
\usepackage{amsfonts}
\usepackage{amsmath}
\usepackage{algorithm}
\usepackage{algorithmic}
\usepackage{graphicx,subfigure,amsmath,amssymb,cite}

\usepackage[dvips]{color}
\usepackage{float}
\ifCLASSINFOpdf
\else
\fi
\newcommand{\Hnull}{\mathcal{H}_0}
\newcommand{\Halt}{\mathcal{H}_1}

\newcommand{\Honull}{\mathcal{{D}}_0}
\newcommand{\Hoalt}{\mathcal{{D}}_1}

\newtheorem{theorem}{\textbf{Theorem}}

\newtheorem{lemma}{\textbf{Lemma}}

\usepackage{cite,graphicx,amsmath,amssymb,cite,algorithm}

\def\BibTeX{{\rm B\kern-.05em{\sc i\kern-.025em b}\kern-.08em
    T\kern-.1667em\lower.7ex\hbox{E}\kern-.125emX}}
\begin{document}

\title{Covert Communication with A Full-Duplex Receiver Based on Channel Distribution Information
}

\author{
\IEEEauthorblockN{Tingzhen Xu$^{1}$, Ling Xu$^{1}$, Xiaoyu Liu$^{1}$, Zaoyu Lu$^{1}$}\\
\IEEEauthorblockA{$^{1}$School of Electronic and Optical Engineering, Nanjing University of Science and Technology, Nanjing, Jiangsu, China}\\
\IEEEauthorblockA{Emails:\{tingzhen.xu, xuling, xiaoyu.liu\}@njust.edu.cn, Lu\_zaoyu@163.com}
}

\maketitle

\begin{abstract}
In this work, we consider a system of covert communication with the aid of a full-duplex (FD) receiver to enhance the performance in a more realistic scenario, i.e., only the channel distribution information (CDI) rather than channel state information (CSI) is known to a warden. Our work shows that transmitting random AN can improve the covert communication with the infinite blocklength. Specifically, we jointly design the optimal transmit power and AN power by minimizing the outage probability at Bob, and we find that the outage probability decreases and then increases as the maximum allowable AN power increases. Intuitively, once AN exceeds an optimal value, the performance will become worse because of the self-interference. The simulation results also show that the performance behaviors of CDI and CSI are different. When Willie only knows CDI, there is an optimal AN power that minimizes Bob's outage probability. However, when Willie knows CSI, the outage probability monotonically decreases with AN power.
\end{abstract}


\section{Introduction}
Privacy and security of information are vital in wireless communications since the information is transmitted through public links\cite{hu2017artificial,shu2018secure,Hu2018covertrelay}. The most common method to ensure the security of information is protecting the content of messages like the traditional encryption technology\cite{encryption}. However, sometimes exposing the location information can be deadly. For example, the commander's position cannot be known to the enemy in a war, otherwise the army will lose a leader. So hiding the existence of the transmitter or communication has become significant. Such scenarios require covert communication to ensure the low probability of being detected by the warden.

The fundamental limits of covert communication in different channels, called  the square root law (SRL),  have been proved in \cite{Bash2013limits}.
The SRL gives the number of bits that $n$ channels can transmit.
Besides the analysis of infinite blocklength, the effect of finite blocklength on covert communications was investigated in \cite{Shihao2018Delay,Yan2017finite}.

A full-duplex (FD) receiver can help achieve a better performance of covert communication compared with adding an external jammer for the fact that a system with an external jammer may cause several issues, such as mobility\cite{Sobers2017uninformedjammer}.
The most works about covert communication are under the premise that the channel state information (CSI) of all channels are known to the transmitter.
However, in practice, the warden cannot exactly know the channel information. For the reasons above,
this work considers a system with a FD receiver to transmit the artificial noise (AN) to the warden Willie in a more realistic scenario (i.e., only the channel distribution information (CDI) is known to Willie)\cite{Lee2014} and draws some conclusions.

\section{System Model}
\begin{figure}[htbp]
\centerline{\includegraphics[width=0.36\textwidth]{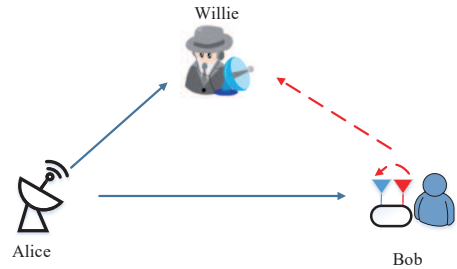}}
\caption{Full-duplex system model.}
\label{system}
\end{figure}

We consider a system of covert communication in Rayleigh fading channel. System model is shown in Fig.~\ref{system}. Alice (a) tries to communicate with Bob (b) covertly and reliably under the supervision of Willie (w), who has to decide whether Alice is transmitting or not. We assume that Alice and Willie are equipped with a single antenna each, while Bob is equipped with two antennas, with one receiving messages and another transmitting AN in order to disturb Willie's receive power. We assume that the average symbol transmit power of Bob $P_b$ is subject to a simple distribution, i.e., the uniform distribution. More complex distributions of $P_b$ will be discussed in future studies. The probability density function (PDF) of $P_b$ is given by.

\begin{equation}\label{Pb}
{f_{{P_b}(y)}} =\left\{
\begin{aligned}
  &\frac{1}{{P_b^{\max }}},&0 \leq {P_b} \leq P_b^{\max }, \\
&0,&\textmd{otherwise},
\end{aligned}
\right.
\end{equation}
where $P_b^{\max }$ is the maximum allowable AN power.

In Rayleigh channel, we consider a more realistic situation. Assume that Willie only knows the CDI of Alice-Willie channel, but we still assume that Alice and Bob know CSI of Alice-Bob channel. The receive signal of Willie is given by
\begin{equation}\label{yw}
{\mathbf{y}_w}[i] =\left\{
\begin{aligned}
&\sqrt {\frac{{{P_b}}}{{r_{bw}^\alpha }}}{h_{bw}} {\mathbf{v}_b}[i] + {\mathbf{n}_w}[i],&{\Hnull},\\
&\sqrt {\frac{{{P_a}}}{{r_{aw}^\alpha }}}{h_{aw}} \mathbf{x}[i] + \sqrt {\frac{{{P_b}}}{{r_{bw}^\alpha }}}{h_{bw}}{\mathbf{v}_b}[i] + {\mathbf{n}_w}[i],&{\Halt}.
\end{aligned}
\right.
\end{equation}
where $P_a$ and $P_b$ are the transmit power at Alice and Bob, respectively, ${h_{j}}$ represents the channel, $j$ can be $ab,aw,bw,bb$. The mean value of ${|{h_j} |}^2$ is ${\text{E}}\left[ {{{\left| {{h_j}} \right|}^2}} \right]={\lambda _j}$. $\mathbf{x}_a$ is the transmit signal at Alice satisfying $\mathbb{E}[\mathbf{x}_a[i]\mathbf{x}^{\dag}_a[i]]=1$, $i = 1, 2, \dots, N$, where $i$ is the index of channel uses, $\mathbf{v}_b$ is the AN transmitted by Bob satisfying $\mathbb{E}[\mathbf{v}_b[i]\mathbf{v}^{\dag}_b[i]]=1$, and $\mathbf{n}_w[i]\thicksim\mathcal{CN}(0,\sigma^2_w)$ is the AWGN at Willie with variance $\sigma^2_w$, $r_{aw}^{\alpha}$ and $r_{bw}^{\alpha}$ representing the distances of Alice-Willie and Bob-Willie, respectively, where $\alpha$ is the path loss exponent. When the number of channel uses $N\rightarrow\infty$, the average symbol received power is
\begin{equation}\label{Tw}
T_w=\left\{
\begin{aligned}
&\frac{{{P_b}}}{{r_{bw}^\alpha }}{|h_{bw}|}^2 + \sigma _w^2,&\Hnull,\\
&\frac{{{P_a}}}{{r_{aw}^\alpha }}{|h_{aw}|}^2 + \frac{{{P_b}}}{{r_{bw}^\alpha }}{|h_{bw}|}^2 + \sigma _w^2,&\Halt.
\end{aligned}
\right.
\end{equation}
Willie has to make a decision between $\Hnull$ and $\Halt$, where null hypothesis $\Hnull$ means Alice does not transmit, the alternate hypothesis $\Halt$ means Alice does transmit covert message to Bob.
\section{Detection Performance of Willie}
We calculate the the false alarm rate (FAR) $\mathbb{P}_{FA}\triangleq \mathbb{P}(\Hoalt|\Hnull)$ and miss detection rate (MDR) $\mathbb{P}_{MD}\triangleq \mathbb{P}(\Honull|\Halt)$ as the metrics to measure the performance of Willie, where $\Honull$ and $\Hoalt$ are decisions made by Willie in favor of $\Hnull$ and $\Halt$, respectively.
\begin{lemma}
The FAR and MDR at Willie are given by

\begin{equation}\label{Pfa_all}
P_{FA}=
\begin{cases}
1,&\tau\leq \rho_1,  \\
P_1,&\tau>\rho_1£¬
\end{cases}
\end{equation}
\begin{equation} \label{Pmd_all}
P_{MD}=
\begin{cases}
0,&\tau\leq {\rho_2},\\
P_2,&\tau>{\rho_2}
\end{cases}
\end{equation}
where $P_1$ and $P_2$ are given by \eqref{Pfa} and \eqref{Pmd}, respectively.
\begin{align} \label{rho}
&\rho _1=\sigma_w^2+\frac{\lambda_{bw}P_b^{\max}}{Nr_{bw}^\alpha}, \\\nonumber
&\rho _2=\sigma _w^2 +\frac{P_a \lambda_{aw}}{Nr_{aw}^\alpha}.   \\\nonumber
\end{align}
\end{lemma}

\begin{IEEEproof}
\begin{align}\label{Pfa}
{P_{FA}}&={P_r}\left( {\frac{{{P_b}}}{{r_{bw}^\alpha }}{{\left| {{h_{bw}}} \right|}^2} + \sigma _w^2 > \tau |{\Hnull}} \right)\\\nonumber
&=\int_{{0}}^{ P_b^{\max} }{\frac{1}{{P_b^{\max }}}} \int_{\frac{(\tau-\sigma_w^2)r_{bw}^\alpha}{y}}^{+\infty}{{\frac{1}{\lambda _{bw}}}{e^{ - {\frac{1}{\lambda _{bw}}}z}}\text{dz}} {\text{dy}}\\\nonumber
&=e^\gamma-\gamma \text{Ei}(\gamma)\\\nonumber
&\triangleq P_1,
\end{align}
where $\gamma=\frac{(\sigma_w^2-\tau) r_{bw}^{\alpha}}{\lambda_{bw}P_b^{\max} }$,
$\text{Ei}(x)=\int_{ - \infty }^x {\frac{{{e^t}}}{t}} \textmd{dt}$ is the exponential integral function.
\begin{align}\label{Pmd}
{P_{MD}}=&{P_r}\left( {\frac{{{P_a}}}{{r_{aw}^\alpha }}{{\left| {{h_{aw}}} \right|}^2} + \frac{{{P_b}}}{{r_{bw}^\alpha }}{{\left| {{h_{bw}}} \right|}^2} + \sigma _w^2 < \tau |{\Halt}} \right)\\\nonumber
=&\int_{{0}}^{ P_b^{\max} }{\frac{1}{{P_b^{\max }}}}\text{dy}
\int_{{0}}^{ + \infty } {\frac{1}{\lambda _{bw}}{e^{ - {\frac{1}{\lambda _{bw}}x}}\text{dx}}}\\\nonumber
&\int_{{0}}^{(\tau-\sigma_w^2-\frac{yx}{r_{bw}^\alpha})\frac{r_{aw}^\alpha}{P_a} } {\frac{1}{\lambda _{aw}}{e^{ - {\frac{1}{\lambda _{aw}}}z}}\text{dz}}\\\nonumber
=& 1-\frac{d}{g}e^{(\sigma_w^2-\tau)g}\ln(\frac{d}{d-g})\\\nonumber
\triangleq & P_2,
\end{align}
where
\begin{equation}\label{d_and_g}
d=\frac{r_{bw}^\alpha}{\lambda_{bw}P_b^{\max}},
~~~g=\frac{r_{aw}^\alpha}{\lambda_{aw}P_a},
\end{equation}
with constraint
\begin{equation}\label{d>g}
d>g. \nonumber
\end{equation}
The derivation of $\rho_1\sim\rho_4$ please refer to Appendix~\ref{proof_rho}
\end{IEEEproof}

\begin{theorem}
Consider the model and metrics above, the detection error probability at Willie is given by
\begin{equation}\label{xi2}
{\xi}=
\begin{cases}
1,&\tau\leq\rho_1,\\
P_1,&\rho_1<\tau\leq\rho_2,\\
P_1+P_2,&\tau>\rho_2,\\
\end{cases}
\end{equation}
The optimal threshold for Willie's detection is
\begin{equation}
\tau^{\ast}=[\rho_2,+\infty).
\end{equation}
\end{theorem}

\begin{IEEEproof}
As per \eqref{rho}, in order to represent  $\xi$, we need to discuss the relationship between $\rho_1$ and $\rho_2$.
When $\rho_1>\rho_2$, we can easily get $\xi=1+P_2$ in interval $[\rho_2,\rho_1]$. Thus the situation of $\rho_1>\rho_2$ is impossible.
For $\rho_2\geq\rho_1$, we have the expression as shown in \eqref{xi2}.

For $\rho_1<\tau\leq\rho_2$, $\xi$ decreases with $\tau$.  When  $\tau>\rho_2$, the first derivative of $\xi$ with respect to $\tau$ is
\begin{equation}\label{xi'}
\xi_{\tau}'=d\text {Ei}(d(\sigma_w^2-\tau))+d\ln(\frac{d}{d-g})e^{g(\sigma_w^2-\tau)}.
\end{equation}
After some calculations and analysis, we can find $\xi=P_1+P_2$ decreases and then increases as $\tau$ increases, thus the minimum $\xi$ is in the interval of  $\tau>\rho_2$.
\end{IEEEproof}

\section{Performance of Covert Communciation}

Next, we analyze the performance of Bob. According to \cite{Hu2017fullduplex}, we have the outage probability of Bob given by
\begin{align}\label{minPot}
{P_{out}} = 1 - {\lambda _{ab}}{e^{ - \frac{{\mu \sigma _b^2}}{{{\lambda _{ab}}}}}}\frac{{\ln \left( {\mu {h}{\lambda _{bb}}P_b^{\max }+\lambda _{ab}} \right) - \ln \left( {{\lambda _{ab}}} \right)}}{{\mu {h}{\lambda _{bb}}P_b^{\max }}}
\end{align}
where $\mu  = \frac{{r_{ab}^\alpha }}{{{P_a}}}\left( {{2^R} - 1} \right)$, $h$ is the self-interference cancellation coefficient at Bob, $R$ is the transmission rate from Alice to Bob.
The problem that minimizes the outage probability $P_{out}$ under the covert constraint is given by
\begin{align}\label{P1}
(\mathbf{P1}) \quad &\min
\limits_{P_a,P_b^{\max}}
P_{out}\\
&\textmd{s.t.}\quad \xi^\ddagger\geq1-\epsilon ,\label{covert_condition}
\end{align}
where $\xi^\ddagger=P_1+P_2$. Here we use $\xi^\ddagger$ for the fact that the minimum $\xi$ is in $[\rho_2,\infty)$.

\begin{theorem}
For any given covert constraint $\epsilon$, the optimal $P_a$, for fixed $P_b^{\max}$, to minimize the effective covert rate is
\begin{align}
P_a^{\dagger}=\frac{r_{aw}^\alpha}{g_2\lambda_{aw}},
\end{align}
And the globally optimal $P_b^{\max}$ is
\begin{align}\label{Pbmax_op}
P_b^{\max*}=\arg \min \limits_{P_b^{\max}}P_{out}^\dagger.
\end{align}
where $g_2$ is given in \eqref{g12}, $P_{out}^\dagger$ is explained later.
\end{theorem}

\begin{IEEEproof}
 $P_{out}^\dagger$ decreases with $P_a$, and $\xi^\ddagger$ first increases then decreases as $P_a$ increases. We omitted the specific proof due to the limited length of paper. Thus we have the optimal $P_a$ is the solution to $\xi^\ddagger=1-\epsilon$.
We define
\begin{align}
M=(\sigma_w^2-\tau)d\text {Ei}((\sigma_w^2-\tau)d)-e^{(\sigma_w^2-\tau)d}-\epsilon,
\end{align}
then the equation $\xi^\ddagger=1-\epsilon$ becomes
\begin{align}
-\frac{d}{g}e^{(\sigma_w^2-\tau)g}\ln(\frac{d}{d-g})=M,
\end{align}
For convenience of calculation and analysis, we approximate the exponential function and the logarithm function with the second and first order Taylor expansion, respectively. Then
\begin{align}\label{quadratic}
-\frac{d}{g}\big[1+(\sigma_w^2-\tau)g+\frac{g^2(\sigma_w^2-\tau)^2}{2}\big]\times\frac{d}{d-g}=M
\end{align}
with constraint $2P_b^{\max}\lambda_{bw}r_{aw}^\alpha<P_a\lambda_{aw}r_{bw}^\alpha$.
To ensure the equation have two positive real roots, it must satisfy
\begin{equation}
\begin{cases}
\Delta\geq0,\\
g_1+g_2>0,\\
g_1\times g_2>0,
\end{cases}
\end{equation}
where
\begin{align}
\Delta =M^2-d^2(\tau-\sigma_w^2)^2+M(4-2d\tau+2d\sigma_w^2).
\end{align}
$g_1$, $g_2$ are the corresponding two roots.
That is
\begin{equation}
\begin{cases}
M<\min \{\frac{1}{2}d^2(\tau-\sigma_w^2)^2,d(\tau-\sigma_w^2)^2\},\\
M^2-d^2(\tau-\sigma_w^2)^2+M(4-2d\tau+2d\sigma_w^2)\geq0.\\
\end{cases}
\end{equation}
Solve quadratic equation of \eqref{quadratic}, we get
\begin{align}\label{g12}
g_{1,2}=\frac{d^2(\tau-\sigma_w^2)^2-dM\pm \sqrt{\Delta}}{d^2\sigma_w^2(\sigma_w^2-2\tau)+d^2\tau^2-2M},
\end{align}

 As per \eqref{d_and_g}, we have $P_a^1$, $P_a^2$ and $P_a^1<P_a^2$. To achieve a larger $P_a$, we keep  $P_a^2$ and get the optimal $P_a$ for fixed $P_b^{\max}$ is $P_a^{\dagger}=P_a^2$.
Substituting $P_a^{\dagger}$ into $P_{out}$, we will obtain $P_{out}^{\dagger}$. Then $P_b^{\max*}$ is derived from \eqref{Pbmax_op}.
\end{IEEEproof}

\section{Simulations and Discussions}\label{AA}
In this section, we evaluate the performance of the whole analysis. For simplicity, we suppose all the distcance $r_j=1$ and $\lambda_j=1$ unless there is a special statement.

\begin{figure}[htbp]
\centerline{\includegraphics[width=0.36\textwidth]{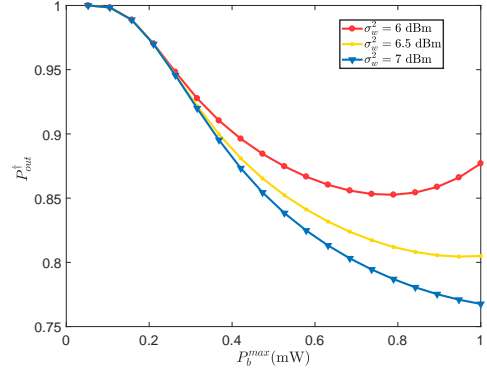}}
\caption{$\overline{R}_c^\dagger$ vs. $P_b^{\max}$, where $\tau=2$, $\epsilon=0.2$, $\lambda_{bw}=0.8$.}
\label{fig:2}
\end{figure}

\begin{figure}[htbp]
\centerline{\includegraphics[width=0.36\textwidth]{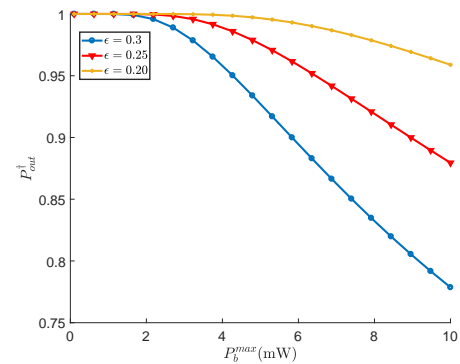}}
\caption{$P_{out}^\dagger$ vs. $P_b^{\max}$ under CDI and CSI, where $\sigma_w^2=\sigma_b^2=0$ dBm, $\epsilon=0.3$, $R=1$, $h=0.1$, $\tau=3$, $\lambda_{bw}=0.5$.}
\label{fig:3}
\end{figure}

Fig.~\ref{fig:2} shows the curves of $P_{out}^\dagger$ versus $P_b^{\max}$ when Willie only knows CDI of Alice-Willie channel. It indicates that there indeed exits an optimal $P_b^{\max}$ that minimizes the outage probability of Bob under the covert constraint. We observe that $P_{out}^\dagger$ first decreases then increases with $P_b^{\max}$ since the AN is helpful for the transmission at first, but when the AN is too large, the performance will be worse due to the self-interference. It also shows that $P_b^{\max\ast}$ increases as $\sigma_w^2$ increases due to the fact that increasing $\sigma_w^2$ means Alice can transmit more power which leads to the larger $P_b^{\max*}$ to cover the transmission.

To show the differences between the cases that Willie only knows CDI and CSI, we plot Fig.~\ref{fig:3}, $P_{out}^\dagger$ versus $P_b^{\max}$ for Willie knows CSI.
 From Fig.~\ref{fig:3}, we know that $P_{out}^\dagger$ monotonically decreases with $P_b^{\max}$ for the fact that the increasing of $P_b^{\max}$ will only benefits Bob's performance when there is no uncertainty at Willie. At the same time, $P_{out}^\dagger$ decreases with $\epsilon$ because that the larger $\xi$ at Willie means the better performance at Bob. By comparing the two figures, we can easily known that under the same conditions, the outage probability at Bob is smaller in the scenario that Willie only knows CDI than the scenario that Willie knows CSI. It means that the former is more beneficial than the latter.

\section{Conclusions}
In this work, we mainly analyze the performance of covert communication under CDI with the aid of a FD receiver. We jointly design $P_a$ and $P_b^{\max}$ to minimize the outage probability. We first optimize $P_a$ by fixing $P_b^{\max}$ and then substitute this $P_a$ into the objective function to obtain $P_b^{\max*}$. Our work shows that transmitting random AN can help improve the covert performance in the case of infinite blocklength.
We have found that $P_b^{\max*}$ increases with $\sigma_w^2$, i.e., if the receive noise variance $\sigma_w^2$ at Wille increases, then  Alice can increase its transmit power $P_a$. This will require that the AN power transmitted by Bob should increase accordingly. Besides, the performance of CDI and CSI is compared. It's consistent with our intuition, i.e., the scenario that Willie only knows CDI performs better than the scenario that Willie knows CSI for covert transmission.
\appendices

\section{derivation of$\rho_1, \rho_2$}
\label{proof_rho}
Apparently, when Willie's detection threshold is smaller than the minimum power of interference-plus-noise, the $P_{FA}$ equals to one according to the definition of false alarm rate, i.e.,
\begin{align}
\tau\leq\sigma_w^2+\frac{P_b^{max}(\min(|h_{bw}|^2))}
{r_{bw}^\alpha}
\end{align}
We set
\begin{align}
Z=\min\{|h_1|^2,|h_2|^2,...|h_N|^2\}
\end{align}
where random variable $X=\{|h_i|^2,1\leq i \leq N\}$, each $|h_i|^2$  is independent and identically distributed. $X\sim E(\lambda)$ and $\lambda=\frac{1}{\lambda_{bw}}$. So we have
\begin{equation}
F_X(x)=\left\{
\begin{aligned}
&0,&x<0,\\
&1-e^{-\lambda x},&x\geq0.
\end{aligned}
\right.
\end{equation}
Then we get
\begin{align}
F_Z(z)
&=P(Z\leq z) \\\nonumber
&=1-P(Z\geq z) \\\nonumber
&=1-P(|h_1|^2\geq z)...P(|h_N|^2\geq z) \\\nonumber
&=1-[1-P(|h_1|^2\leq z)]...[1-P(|h_N|^2\leq z)] \\\nonumber
&=1-[e^{-\lambda z}]^N
\end{align}
So the PDF is
\begin{align}
&f_Z(z)=N\lambda (e^{-\lambda z})^N
\end{align}
The expectation is
\begin{align}
E(z)&=\int_{{0}}^{ +\infty }{zN\lambda (e^{-\lambda z})^N}\text{dz} \\\nonumber
&=\frac{1}{\lambda N}
\end{align}
Finally, $\rho_1$ is derived. And the derivation of $\rho_2$ is similar to $\rho_1$.

\bibliographystyle{IEEEtran}
\bibliography{IEEEabrv,CC}

\end{document}